\documentstyle[prl,aps,twocolumn,psfig]{revtex}

\begin{document}

\twocolumn[\hsize\textwidth\columnwidth\hsize\csname
@twocolumnfalse\endcsname

\draft

\title{Scatterer that leaves "footprints" but no "fingerprints"}
\author{Er'el Granot \footnotemark}
\address{Department of Electrical Engineering, College of Judea and Samaria, Ariel 44837, Israel}

\date{\today}
\maketitle
\begin{abstract}
\begin{quote}
\parbox{16 cm}{\small
We calculate the exact transmission coefficient of a quantum wire
in the presence of a single point defect at the wire's cut-off
frequencies. We show that while the conductance pattern (i.e., the
scattering) is strongly affected by the presence of the defect,
the pattern is totally \emph{independent} of the defect's
characteristics (i.e., the defect that caused the scattering
cannot be identified from that pattern).}
\end{quote}
\end{abstract}


]

\narrowtext \footnotetext{erel@yosh.ac.il} \noindent One of the
most common ways to investigate the inner structure of a system is
to perform scattering studies. That is, by looking at the
scattering {\em pattern} one can cull some notion about the
scatterer that was the cause of the specific scattering pattern.
Our experience shows that every scatterer has a different
scattering pattern. That explains the ubiquity of scattering
techniques in the diagnostic world: for crystallographic studies,
x-rays are used; visible light is usually used to detect molecule
energy levels; ultrasound waves are commonly used for embryo
imaging, etc.

\bigskip
In this paper, we discuss a case of a narrow wire in which our
experience (that every scatterer has a different scattering
pattern) fails. In this case, the scatterer has a strong influence
on the dynamics of the system, both in terms of conductance
(either high or low) and on the conduction pattern. However, the
conductance and the scattering pattern are {\em totally
independent} of the scatterer. The scatterer's elusive conduct can
be phrased: {\em One can see the scatterer's "footprints" (its
strong influence), but cannot see its "fingerprints" (anything
that may assist to characterize it).}

\bigskip
When we think of a small and weak scatterer, the thing we usually
have in mind is a scatterer whose influence on scattering is
negligible. One of the reasons for this is that we are accustomed
to a 3D world. In this case, the cross section is $\sigma \sim
V^{2}$ (see ref.\cite{Merzbacher_70}), where $V$ is the scatterer
strength (potential), i.e., it vanishes with scatterer potential.
In 1D, however, this is definitely not the case. It is well known
that when the incident particles' energy is considerably lower
(see below) than the scatterer's strength (i.e., the scatterer's
potential), most of the incident particles are reflected from the
scatterer, i.e., the scattering is strong regardless of scatterer
"weakness" (so long, of course, as the particles' energy is lower
than the scatterer's potential). This behavior can be presented
easily in the case of a weak scatterer, in which the reflection
coefficient is related to the 1D scattering cross section. By
using the term "weak scatterer" we refer to the case in which its
strength (its weak potential $\Delta V$) and its width ($L$)
satisfy $\sqrt{\Delta V}L\ll1$ (with the units $\hbar=2m=1$). In
this case, the reflection coefficient maintains
\begin{equation}
R\simeq \frac{1}{1+4\omega /\left( \Delta VL\right) ^{2}}
\label{weak_sc}
\end{equation}

where $\omega $ is the energy of the incident particles. One can
easily be convinced, though quite surprisingly, that the extreme
case of the \emph{infinitely shallow} potential barrier is
actually the 1D delta function. That is, for the potential barrier
$\alpha \delta \left( x\right) $ (or, the limit of $\Delta
V=\alpha /L$ for $L\rightarrow 0$), the reflection coefficient
reads

\begin{equation}
R=\frac{1}{1+4\omega /\alpha ^{2}}  \label{1D_delta}
\end{equation}

(notice, that now this is an equation and not merely an
approximation). Eqs. (\ref{weak_sc}) and (\ref{1D_delta}) despite
their simplicity, hold some peculiarities, which cannot be found
in 3D scattering. These peculiarities can be summarized in three
points:

1. The scattering is \emph{increased} when the energy decreases.

2. The scattering is strong despite the scatterer's ''weakness''.

3. The scattering for $\omega \rightarrow 0$ is \emph{independent}
of the scatterer (it does not depend on $\alpha$).

\bigskip

The third point is probably the most peculiar, since it
contradicts our statement that each scatterer has a distinct
scattering pattern. However, in 1D this feature is hardly
interesting since it is valid only for zero incident particle
energy ($\omega =0$). A particle with zero energy has little
chance of even reaching the scatterer. In quasi-1D systems, the
situation can be quite different.

In the case of the thin wire, for example, there are an infinite
number of threshold (cut-off) energies. When the incident
particles have exactly the cut-off energy of the $m$th mode, no
energy is transferred to it (to the $m$th mode), since the
momentum (or the k-vector) of this mode in the propagation
direction is zero. Therefore, it makes sense to expect to find all
the peculiarities of the 1D case, even in a 2D wire, near the
threshold energies.

For such a system (point impurity in a quasi-1D wire) the 2D
Scr\"{o}dinger equation is

\begin{equation}
\nabla^{2}\psi+\left( \omega-V
\right)\psi=-D\left(\mathbf{r}-\mathbf{r}_0 \right) \psi
\label{schr_eq}
\end{equation}

(where we use the units $\hbar=2m=1$). $V$ is the potential of the
wire walls ($V=0$ inside the wire and $V=\infty$ outside it), $D$
is the defect potential and $\mathbf{r}_0=\varepsilon \hat{y} $ is
the impurity location (see Fig.1). Since the defect has the
proprties of a point-like impurity, the right-hand term of the
Scr\"{o}dinger equation can be written $D \left(
\mathbf{r}-\mathbf{r}_0 \right)\psi \left(\mathbf{r}_0 \right)$
\cite{Azbel_91}, which allows for an exact scattering solution.

\begin{figure}
\psfig{figure=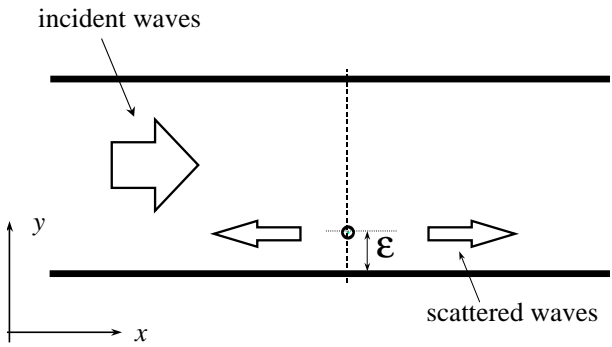,width=8cm,bbllx=160bp,bblly=80bp,bburx=690bp,bbury=470bp,clip=}
\caption{\small \it A 2D wire with a single point defect (black
dot)} \label{fig1}
\end{figure}

Let us assume that we hit the impurity with the incident wave
$\psi_{inc}$. Taking advantage of the point-like nature of the
impurity, the scattered wave function due to the defect is
\cite{Granot_Azbel_94,Granot}

\begin{equation}
\psi _{sc}=\psi _{inc}-\frac{\psi _{inc} \left( \mathbf{r}_{0} \right) \int%
d {\mathbf{r'}} D \left( {\mathbf{r'}}-{\mathbf{r}}_{0}\right) }{%
1+\int d{\mathbf{r'}}G^{+}\left( {\mathbf{r'}},{\mathbf{r}}%
_{0}\right) D\left( \mathbf{r}^{\prime }\mathbf{-r}_{0}\right)
}G^{+}\left( \mathbf{r,r}_{0}\right)  \label{gen_solution}
\end{equation}

where $G^{+}\left( \mathbf{r',r''}\right)$ is the "outgoing"
2D-Green function of the geometry (the wire). It should be noted
that eq. \ref{gen_solution} is an exact solution, however if the
impurity were not an ideal \emph{point} impurity, this equation
would be a first-order approximation in the asymptotic solution
$|\mathbf{r} \rightarrow \infty|$. The Green function for the
given wire geometry takes the form:
\begin{equation}
\begin{array}{l}
  G \left( \mathbf{r,r'} \right)= \\
  i\sum_{n=1}^{\infty}\frac{\sin(n \pi y) %
  \sin(n \pi y')}{\sqrt{\omega - (n \pi)^2}}\exp \left[ i \sqrt{\omega - %
  (n \pi)^2} |x-x'| \right]
\end{array}
  \label{green}
\end{equation}

where ${\mathbf{r}}\equiv x \hat{x}+y \hat{y}$ and ${\mathbf{r'}}
\equiv x' \hat{x}+y' \hat{y}$. Hereinafter, the length parameters
are normalized to the wire's width.

\bigskip

Choosing the right potential for the impurity is a very tricky
business as can be understood from the literature
\cite{Chu_Sorbello_89,Kumar_91,Bagwell_90,Levinson_Lubin_Sukhorukov_92,Gurvitz_Levinson_93,Tekman_90,Kim_99}.
A simple 2D delta function (2DDF), which is a natural candidate to
represent a point impurity (like in 1D), i.e., $\delta \left(
x\right) \delta \left( y\right) $ does not scatter (its cross
section is zero). Throughout this article we us the Impurity D
Function (IDF) that was first presented by Azbel \cite{Azbel_91}.
However, since in our wire's geometry the problem's symmetry is
Cartesian rather than radial, we choose the following IDF:

\begin{equation}
D\left( \mathbf{r}\right) \equiv  \lim_{\rho \rightarrow 0}\frac{2%
\sqrt{\pi }\delta \left( x\right) }{\rho \ln \left( \rho /\rho _{0}\right) }%
\exp \left( -y^{2}/\rho ^{2}\right) .  \label{point_imp}
\end{equation}

Unlike the 2DDF, this potential, which is infinitely shallower
than the 2DDF, does scatter\cite{Azbel_91}. The de-Broglie
wavelength of the impurity's bound state is $\lambda_{B}=\pi
\rho_0 \exp (\gamma /2)/2$ (where $\gamma \simeq 0.577$ is the
Euler constant). This is the only parameter that characterizes the
impurity, and therefore eq. \ref{point_imp} can be used to mimic
any impurity with the same de-Broglie wavelength, where its width
is much smaller than $\lambda_{B}$.

\bigskip
On the face of it, the solution is straightforward: simply to
substitute eqs. \ref{point_imp} and \ref{green} into eq.
\ref{gen_solution}. The problem is that the Green function has a
logarithmic singularity at $|\mathbf{r-r'}|\rightarrow 0$. Here is
where the impurity's width $\rho$ plays a major part, and the
limit $\rho \rightarrow 0$ should be taken with great caution.
Therefore, we first solve the integral for a finite $\rho$ and
only then evaluate the limit.

Let us assume that the incident wave is the $n$th mode, and that
the incident energy is close to the $m$th threshold energy (i.e.,
$\omega \simeq (m \pi)^2$ ) therefore,

\begin{equation}
\psi_{inc} \left( \mathbf{r} \right)=\sin (n \pi y) \exp \left(
i\sqrt{\omega-(n \pi)^2} x \right). \label{incident}
\end{equation}

The probability density of Eq.(\ref{incident}) is presented in
Fig.2 for $n=1$ and $m=2$.

By using the following relation
\begin{equation}
\begin{array}{r}
  \int_{-\infty}^{\infty} dy \sin(n \pi y) \exp \left[
-(y-\varepsilon)^2/\rho^2 \right]=  \\
   \rho \sqrt{\pi} \sin(n \pi \varepsilon) \exp \left[-(n \pi \rho/2)^2 \right]
\\
\end{array}
\label{intlimit}
\end{equation}

We find the solution (for $x>0$)
\begin{equation}
\psi_{sc} \left( \mathbf{r} \right)=\sum_{l=1}^{\infty} \left(
\delta_{nl}-A_{nl} \right) \sin(l \pi y) \exp \left( i
\sqrt{\omega-(l \pi)^2} x \right) \label{Fsolution}
\end{equation}

where $\delta_{nl}$ is the Kronecker delta and

\begin{equation}
A_{nl} \equiv \frac{\sin (n \pi \varepsilon) \sin (l \pi
\varepsilon)}{i \sqrt{\omega-(l \pi)^2} \left[
\frac{\ln(\rho_0/\bar{\rho})}{2 \pi}+\sum_{n'\leq m}
\frac{\sin^2(n' \pi \varepsilon)}{i \sqrt{\omega-(n' \pi)^2}}
\right]} \label{A_def}
\end{equation}

and $\bar{\rho}$ is some length scale which depends on the
impurity's location ($\varepsilon$), the incident energy $\omega$
and $m$:

\begin{equation}
\ln ( \bar{\rho} ) \equiv \lim_{\rho \rightarrow 0} \left\{ \ln
\rho +2 \pi \sum_{n'=m+1}^{\infty} \frac{\sin^2 (n' \pi
\varepsilon)}{\sqrt{(n' \pi)^2-\omega}} e^{-(n' \pi \rho/2)^2}
\right\} \label{rho_bar}
\end{equation}

\begin{figure}
\psfig{figure=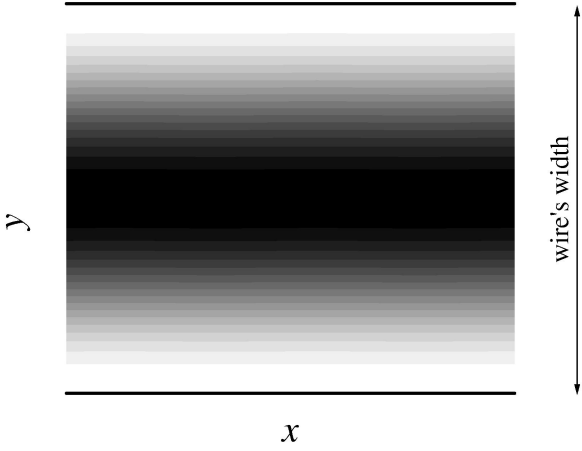,width=8cm,bbllx=70bp,bblly=50bp,bburx=430bp,bbury=360bp,clip=}
\caption{\small \it The probability density distribution in the
wire for $n=1$ and $\omega=(2\pi)^2$ in the absence of the point
defect.} \label{fig2}
\end{figure}

In this paper we discuss the case where $\omega \simeq (m \pi)^2$
for any integer $m$ (though the figures are focused on the case
$\omega \simeq (2 \pi)^2$, i.e., $m=2$). In this particular case
only the $n$th mode (the incident mode) and the $m$th one have a
considerable influence on the scattering

\begin{equation}\label{scatt_m_mode _vicinity}
\psi_{sc} \simeq \psi_{inc}- \frac{\sin(n \pi \varepsilon)}{\sin(m
\pi \varepsilon)} \frac{\sin(m \pi y) e^{i \sqrt{\omega-(m \pi)^2}
|x| }}{1+i \sqrt{\omega-(m \pi)^2/\Delta_m}}
\end{equation}

where

\begin{equation}\label{delta_m_def}
  \Delta_m^{-1/2} \equiv \frac{\ln(\rho_{0}/\bar{\rho})}{2 \pi \sin^2(m \pi
  \varepsilon)}-i \sum_{n'<m} \frac{\sin^2(n' \pi \varepsilon)}{\sin^2(m \pi
  \varepsilon)}\frac{1}{\pi \sqrt{m^2-n'^2}}
\end{equation}


The scattered wave function, i.e., Eq.(\ref{scatt_m_mode
_vicinity}), depends on the
scatterer's parameter ($\rho_0$) only via $\Delta _{m}$%
. Therefore, when
\begin{equation}\label{threshold_vicinity}
(\omega-(m \pi)^2)/\Delta_m \ll 1
\end{equation}

one finds:

1) When the energy is not close to the threshold energy, the
scattering is negligible; as we get closer to the threshold
energy, the scattering increases.

2) The scattering coefficient is large (can have \emph{any} value)
regardless of the scatterer's "weakness".

3) Near the threshold energies, the scattering is
\emph{independent} of the scatterer (it does not depend on the
scatterer's parameter).

\bigskip
Again, the most bizarre behavior is the third one, which is
manifested in the limit $\omega \rightarrow (m \pi)^2$ of
Eq.(\ref{scatt_m_mode _vicinity}) (see Fig.3):

\begin{equation}\label{scatt_m_mode _exact}
\psi_{sc} = \psi_{inc}- \frac{\sin(n \pi \varepsilon)}{\sin(m \pi
\varepsilon)} \sin(m \pi y)
\end{equation}

Eq.(\ref{scatt_m_mode _exact}), which present the scattering of
the $n$th mode, when its energy is equal to the threshold energy
of the $m$th one, can be generalized for a wire with an arbitrary
(but uniform) cross section

\begin{equation}\label{scatt_m_mode _exact_arb}
\psi_{sc} = \chi_n(y) e^{i \sqrt{\omega_m-\omega_n} x}-
\frac{\chi_n(\varepsilon)}{\chi_m(\varepsilon)} \chi_m(y)
\end{equation}

where $\chi_n(y)$ is the transversal eigenstates of the wire, with
the corresponding eigen energies $\omega_n$.

\begin{figure}
\psfig{figure=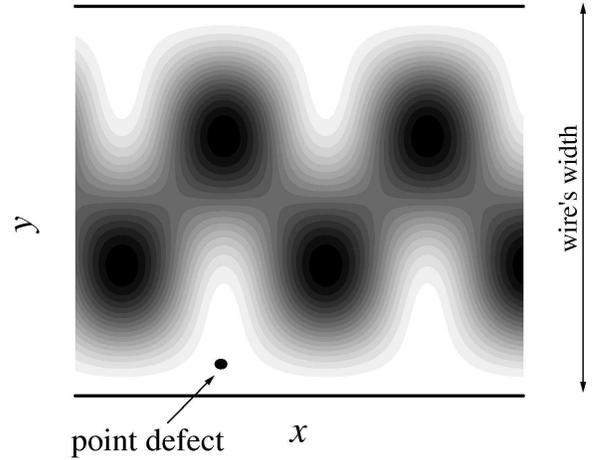,width=8cm,bbllx=70bp,bblly=50bp,bburx=430bp,bbury=360bp,clip=}
\caption{\small \it The probability density distribution in the
wire for $n=1$ and $\omega=(2\pi)^2$ when the point defect is
present.} \label{fig3}
\end{figure}

While Eq.(\ref{scatt_m_mode _vicinity}) is an approximation,
Eqs.(\ref{scatt_m_mode _exact}) and (\ref{scatt_m_mode
_exact_arb}) are totally accurate for \emph{any point impurity},
and at any impurity's location.

In the case of a surface impurity, i.e., $\varepsilon \rightarrow
0 $ (or $\varepsilon \rightarrow 1$), eq. (\ref{scatt_m_mode
_vicinity}) is reduced to an even simpler one

\begin{equation}\label{scatt_m_mode _vic_boundary}
\psi_{sc} \simeq \psi_{inc} \mp \left( \frac{n}{m} \right)
\frac{\sin(m \pi y) e^{i \sqrt{\omega-(m \pi)^2} |x| }}{1+i
\sqrt{\omega-(m \pi)^2/\Delta_m}}
\end{equation}

where

\begin{equation}\label{delta_m_def_bound}
  \Delta_m^{-1/2} \simeq \frac{\ln(\rho_{0}/C \varepsilon)}{2 \pi (m \pi
  \varepsilon)^2}
\end{equation}

and
\begin{equation} \label{C_constant}
C \equiv 4 \exp \left[ \gamma /2 - \mathrm{Ci} (\pi) \right]
\simeq 5
\end{equation}

is a numerical constant (Ci is the cosine integral). The upper
sign (minus) in eq.(\ref{scatt_m_mode _vic_boundary}) stands for
impurity at the lower boundary $\varepsilon \ll m^{-1}$ while the
plus implies an upper boundary impurity $1-\varepsilon \ll m^{-1}$
(in this case the $\varepsilon$ should be replaced by
$1-\varepsilon$ in eq.\ref{delta_m_def_bound}).

Thus, at the threshold energy, i.e., $\omega = (m \pi)^2$,

\begin{equation} \label{scatt_m_mode _exac_boundary}
\psi_{sc} = \sin(m \pi y) e^{i \pi \sqrt{m^2-n^2} x} \mp \left(
\frac{n}{m} \right) \sin(m \pi y)
\end{equation}

That is, in the case of a \emph{surface} impurity then close
enough to the threshold energies (i.e., when eq.(\ref{scatt_m_mode
_exac_boundary}) holds) the scattering is also independent of the
impurity's \emph{location}. Any impurity's characteristics have
faded away near the threshold energies. Eq.(\ref{scatt_m_mode
_exac_boundary}) does not reflect any feature of the scatterer: it
depends neither on its strength nor on its location.

It was shown in the literature (see, for example,
ref.\cite{Chu_Sorbello_89}) that at the threshold energies, the
conductance is totally quantized and is independent of the point
defects, however, here we show two additional results:

the scattering is not a negligible quantity, it does not affect
the conductance but it does distort the conduction \emph{pattern};
and at the same time, that this severe distortion is
\emph{independent of the scatterer that caused it}.

It should be stressed that while the discussion was focused on
quantum wire, this effect can occur in any waveguide with a single
point scatterer: acoustical waveguide, electromagnetic waveguide,
optical waveguide, etc.

I am grateful to Mark Azbel for enlightening discussions.

\end{document}